\documentclass[conference]{IEEEtran}
\IEEEoverridecommandlockouts

\usepackage{subcaption}  
\usepackage{color}
\usepackage{pifont}
\usepackage{bbm}
\usepackage{stfloats}
\usepackage[short]{optidef}
\usepackage{multirow}
\usepackage{comment}

\usepackage{amsmath,amssymb,amsthm}
\usepackage{ragged2e} 

\usepackage{array}
\IEEEoverridecommandlockouts

\usepackage{cite}
\usepackage{amsmath,amssymb,amsfonts}
\usepackage{algorithmic}
\usepackage{graphicx}
\usepackage{textcomp}
\usepackage{xcolor}
\usepackage{amsmath}

\def\BibTeX{{\rm B\kern-.05em{\sc i\kern-.025em b}\kern-.08em
    T\kern-.1667em\lower.7ex\hbox{E}\kern-.125emX}}
\begin{document}
\title{Joint Task Offloading and User Scheduling in 5G MEC under Jamming Attacks 
}

 \author{\IEEEauthorblockN{Mohammadreza Amini, Burak Kantarci, Claude D'Amours, and Melike Erol-Kantarci}\\  \vspace{-0.15in}
 \IEEEauthorblockA{School of Electrical Engineering and Computer Science, University of Ottawa, Ottawa, ON, Canada \\
 \texttt{\{mamini6,burak.kantarci, cdamours, melike.erolkantarci\}@uottawa.ca}}
 \vspace{-0.35in}
 }

\maketitle

\begin{abstract}
In this paper, we propose a novel joint task offloading and user scheduling (JTO-US) framework for 5G mobile edge computing (MEC) systems under security threats from jamming attacks. The goal is to minimize the delay and the ratio of dropped tasks, taking into account both communication and computation delays. The system model includes a 5G network equipped with MEC servers and an adversarial on-off jammer that disrupts communication. The proposed framework optimally schedules tasks and users to minimize the impact of jamming while ensuring that high-priority tasks are processed efficiently. Genetic algorithm (GA) is used to solve the optimization problem, and the results are compared with benchmark methods such as GA without considering jamming effect, Shortest Job First (SJF), and Shortest Deadline First (SDF). The simulation results demonstrate that the proposed JTO-US framework achieves the lowest drop ratio in the presence of the jammer and effectively manages priority tasks, outperforming existing methods. Particularly, when the jamming probability is 0.8, the proposed framework mitigates the jammer's impact by reducing the drop ratio to $63\%$, compared to $89\%$ achieved by the next best method.
\end{abstract}


\section{Introduction} \label{Sec:Intro}

The rapid progress of 5G networks has enabled applications requiring ultra-reliable, low-latency communication (URLLC), massive machine-type communication (mMTC), and enhanced mobile broadband (eMBB) \cite{5Gsurvey}. Mobile edge computing (MEC) plays a key role in meeting these demands by offloading tasks to nearby servers, reducing latency, and improving quality of service \cite{Liu2023,Zhang2024, Moshiri2024}.

5G networks, especially those for critical infrastructure and mission-critical applications, face significant challenges from adversarial threats like jamming attacks \cite{Nencioni_2023, asemian2025active, Lohan.2024, asemian2024dtddnn}. Jammers disrupt communication by transmitting interference signals, leading to increased dropped task rates and reduced service reliability \cite{Rodrigo2018, Arjoune_2020, amini2024bypassing}. These attacks risk the reliability and efficiency of 5G-enabled MEC systems, and call for the need for robust task offloading and user scheduling frameworks to minimize security threats. Despite its importance, there is limited research on mitigating jamming effects in MEC.

Some studies explore using cooperative jammers at the network level to counter eavesdropping. For example, \cite{Lu2022} deploys a ground jammer to disrupt UAV eavesdroppers, enhancing resource allocation and trajectory planning. In \cite{Lu2022_2}, a security scheme for NOMA-based UAV-MEC networks is proposed, maximizing security computation capacity while meeting minimum security requirements for ground users. This study utilizes worst-case scenario analysis and optimization methods like successive convex approximation and block coordinate descent to improve transmit power and UAV trajectories, yielding significantly better security performance. A multi-UAV MEC system is introduced in \cite{Zhang2024_2}, optimizing task offloading, resource allocation, and UAV trajectories with jamming signals from the base station to mitigate eavesdropping risks. The Joint Dynamic Programming and Bidding (JDPB) framework effectively addresses these challenges using Successive Convex Approximation (SCA) and Block Coordinate Descen (BCD) methods, demonstrating superior performance in simulations. Finally, \cite{Yan2024_task} examines an OFDMA-based MEC system in 5G, tackling security risks in uplink signals through physical layer security techniques, including server-induced jamming and co-frequency interference. Their proposed framework minimizes latency while optimizing resource allocation and task offloading policies, ensuring secure communication with lower computational complexity. All these studies highlight the coordinated use of jamming to reduce eavesdropping capabilities.

In the context of security in the RF domain with the presence of a jammer as an attacker, \cite{Xiao_2020} proposes a Reinforcement Learning (RL)-based solution in the task offloading scenario. The aim is to optimize offloading decisions, such as edge device selection, transmit power, and offloading rate by employing a Deep RL (DRL) method using actor and critic networks. However, the approach is not applicable in multi-user scenarios as the decision is made at the local device without considering network traffic. One of the latest works in jamming mitigation in MEC scenario that uses DRL to mitigate the effect of jammer is \cite{Liu2024}. Authors integrate semantic communication with MEC using UAVs to tackle jamming attacks. A DRL approach is proposed to jointly optimize UAV trajectories, user associations, and channel selections, minimizing task completion time and maximizing semantic spectral efficiency (SSE). The approach effectively mitigates the impact of jamming, and simulations show it outperforms baseline methods in terms of SSE and task completion time. These are among a very few papers that pay attention to the task offloading in MEC under the presence of a jammer attacker. In this context, this paper is unique in a sense that it proposes a novel joint task offloading and user scheduling (JTO-US) schema to address both task prioritization, user scheduling, and task offloading under jamming attacks. The proposed method aims to minimize the tasks' delay and dropped task ratio by incorporating the effects of jamming attacks into the scheduling-offloading decision process. Through comprehensive simulations, we demonstrate that the propsed framework significantly outperforms existing approaches in terms of dropped task ratio and latency, even in high jamming probability scenarios. The main contributions of this work are as follows:

\begin{itemize}
    \item Formulate and analyze network metrics such as dropped task ratio and task latency in a multi-user MEC scenario under the presence of an \emph{on-off jammer}.
    \item Develop a joint task offloading and user scheduling strategy aimed at reducing the dropped task ratio and latency, while mitigating the impact of a jammer.
\end{itemize}

The rest of this paper is organized as follows. Section \ref{Sec:Sys_model} describes the system model including network and adversary model. The problem formulation is presented in the Section \ref{Sec:Formulation}. Numerical Results and discussions are included in Section \ref{Sec:Neumerical_results}. Finally, some conclusions are given in Section \ref{Section conclusion}.

\section{System Model  } \label{Sec:Sys_model}





Network model and the behavior of the jammer are presented in this section. We consider a 5G cellular network in which each gNB is equipped with a MEC server for intensive task computations. We further assume that there are $N$ total UEs randomly distributed in the sector area of the cell. Each user is served through a specific beam as multi-user MIMO is assumed. Fig. 1 shows the network model. Let $\mathcal{N}_a$ be the set of active UEs having their task ready to transmit to the MEC server with the number of elements in this set given by $||\mathcal{N}_a||=N_a$, $N_a \leq N$. Transmission protocol follows 5G air interface in which transmission time intervals (TTIs) are divided into frames. Each frame is divided into $S$ mini-slots in time and $R$ resource blocks (RBs) in the frequency domain. At the beginning of each frame, the gNB informs active users of their transmission schedule, i.e., which mini-slot is assigned to each user for task transmission. Each mini-slot duration is $T_{s}=\frac{T_f}{S}$ where $T_f$ is the frame duration.
The MEC server at the gNB has a computational capacity of $C$ cycles per second. 
\graphicspath{{figs/}}
\begin{figure}[htp]
    \centering
    \includegraphics[width=1\linewidth]{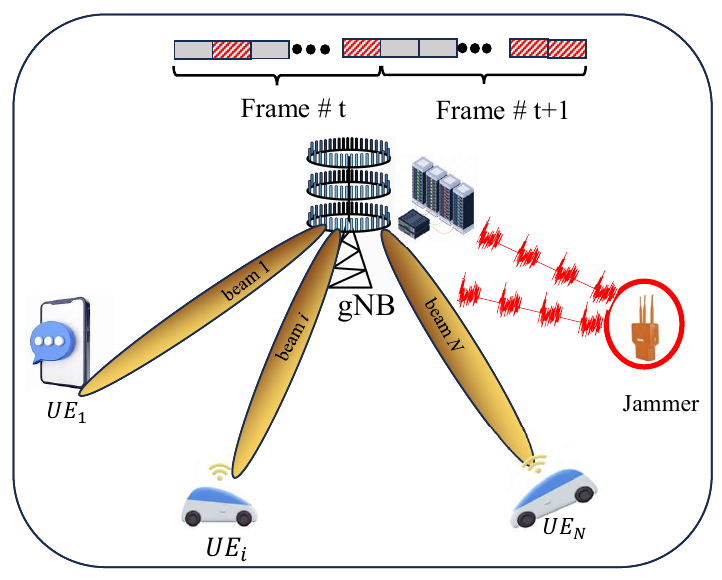}
    \caption{Network Structure.}
    \label{fig: network}
 \end{figure}

The adversary is modeled as a jammer disrupting communication links by transmitting the jamming signal. It is assumed that a jammer is of \emph{on-off} type which means each mini-slot at each frame is jammed with a certain probability. We assume that the \emph{on-off} jammer behavior can be modeled as a \emph{Hidden Discrete Time Markov Chain} (HDTMC) at the gNB. The HDTMC parameters can be estimated through some learning process by observing the environment, as is done in \cite{Huda2014, ARCHER2002}. Therefore, we denote the jamming probability of the $j^{th}$ mini-slot in the $t^{th}$ frame for the $i^{th}$ user\footnote{Note that due to beamforming at the gNB, each user signal experiences different jamming power at the destination.} by $P^{t}_{i,j}$. Due to the presence of the jammer, some tasks may be left unassigned to any slots and are therefore considered dropped. Additionally, tasks that the gNB scheduler cannot process before their deadlines are also treated as dropped.

\section{Problem Formulation} \label{Sec:Formulation}

The aim of the joint task offloading and scheduling framework is to schedule tasks in the server and users in the communication slots so that the transmission and processing of their tasks experience minimum delay and drop task ratio in the presence of the jammer. To this aim, task offloading at the server and user scheduling on air are formulated in this section. 

Let $X^t_{i,j} \in \{0,1\}$ be a user scheduling indicator in which $X^t_{i,j}=1$ means $i^{th}$ user is scheduled at the $j^{th}$ mini-slot of the $t^{th}$ frame for data transmission. Each task -- say $T_i$\footnote{It is assumed that each user has only one task at a single frame.}-- is characterized by the tuple $(c_i,t^i_d,w_i)$ in which $c_i$ is the number of task cycles, $t^i_d$ is the task deadline, $w_i$ is the task weight, indicating how urgent (level of importance) is the task if dropped. Then, the processing time of the $i^{th}$ task at the server equals $t_p^i=\frac{c_i}{C}$ where $C$ is the processing speed of the server in cycle per second.

It is not possible to assign more than one task to one mini-slot and some mini-slots might left without any task due to the scheduler decision. Hence, we have (\ref{eq: X_condition_1}) as, \vspace{-4mm}

\begin{equation} \label{eq: X_condition_1}
\begin{split}
    \sum_{i=1}^{N_a} X^t_{i,j} \leq 1 , \quad \forall j=1, ..., S 
   \end{split}    
\end{equation}
\vspace{-4mm}

Furthermore, the total number of assigned mini-slots must not exceed the number of active users. Therefore, there is another constraint as (\ref{eq: X_condition_2}). \vspace{-4mm}

\begin{equation} \label{eq: X_condition_2}
\begin{split}
        \sum_{i=1}^{N_a}\sum_{j=1}^{S} X^t_{i,j} \leq N_a  
\end{split}    
\end{equation}

To address the task offloading problem, an additional indicator is introduced. Let $Y^t_{i,k} \in \{0,1\}$ be a binary variable that indicates the position of task $T_i$ in the server queue at the $t^{th}$ frame, where $Y^t_{i,k}=1$ signifies that $T_i$ is the $k^{th}$ task to be processed by the MEC server. Latency and dropped task ratio are the main metrics considered in task offloading problems. 


To calculate the latency, some preliminary metrics are calculated. Assuming the $i^{th}$ task is scheduled during the $j^{th}$ mini-slot, i.e., $X^t_{i,j}=1$, its position in the server queue at gNB is calculated in (3) as, \vspace{-4mm}

\begin{equation} \label{eq: Y_i_k}
    \begin{split}
        Y^t_{i,k}=\begin{cases}
            1, \quad \textit{if } \, k=\sum_{i=1}^{N_a}\sum_{j=1}^{J} X^t_{i,j}    \textit{  and  } X^t_{i,J}=1  \\
            0, \textit{  if  } \sum_{j=1}^{S} X^t_{i,j}=0, \, \forall k
        \end{cases}
    \end{split}
\end{equation}

The above equation formulates if no mini-slots is assigned to $T_i$, it is dropped and hence, no position is allocated in the server. However, if it is assigned to the $J^{th}$ mini-slot, only the assigned mini-slots up to $J^{th}$ one is considered in the position of $T_i$ in the server queue. This way, the scheduler attempts to minimize both transmission delay and waiting time in the queue. Note that for all tasks we have $k \leq J$. However, there is no need to add it as an additional constraint since it is considered in (\ref{eq: Y_i_k}). Moreover, the total number of tasks in the server queue can be written as, \vspace{-4mm}

\begin{equation}  \label{eq: total_task_queue}
    \begin{split}
      N_q=\sum_{i=1}^{N_a}\sum_{j=1}^{S} X^t_{i,j} 
    \end{split}
\end{equation} 
It is worth noting that for all tasks $k \leq N_q$. Nevertheless, we do not need to consider this constraint for the same reason mentioned.
Therefore, the position of $T_i$ in the server queue is derived as, \vspace{-4mm}

\begin{equation}  \label{eq: T_i_position}
    \begin{split}
      K_i=\Bigg(\sum_{j=1}^{S}  X^t_{i,j}\Bigg)  \sum_{k=1}^{N_q} k Y^t_{i,k} 
    \end{split}
\end{equation} 
Note that if there is no mini-slot assigned to $T_i$, then $K_i=0$.

To simplify (\ref{eq: Y_i_k}), \emph{Sign} function is used to write it in one line. To do this, we note that the mini-slot assigned to the $i^{th}$ user can be obtained as,\vspace{-5mm}

\begin{equation}  \label{eq: T_i_slot}
    \begin{split}
      J=\sum_{j=1}^{S} j X^t_{i,j} 
    \end{split}
\end{equation} 
If no mini-slot is assigned to $T_i$, then $J=0$, otherwise $J \neq 0$. Therefore, (\ref{eq: Y_i_k}) can be written in one line based on \emph{Sign} and \emph{Dirac delta} functions as per (7), \vspace{-4mm}

\begin{equation}  \label{eq: T_i_slot}
    \begin{split}
      Y^t_{i,k}=Sign(J) \, \,  \delta \Big(k- \sum_{i=1}^{N_a}\sum_{j=1}^{J} JX^t_{i,j}\Big)
    \end{split}
\end{equation}

As mentioned, task latency, $\mathcal{L}_i$ is a very important metric in task offloading scenarios. It consists of two parts namely, communication delay, $\mathcal{L}_{i,c}$, and waiting time in the queue, $\mathcal{L}_{i,q}$. Considering the mini-slots assigned to $T_i$, its communication latency can be easily written as, \vspace{-4mm}

\begin{equation}  \label{eq: comm_latency}
    \begin{split}
      \mathcal{L}_{i,c}= \sum_{j=1}^S j T_{s} X^t_{i,j}
    \end{split}
\end{equation}

Then, the total communication delay of all UEs is,

\begin{equation}  \label{eq: sum_comm_latency}
    \begin{split}
      \mathcal{L}_{c}= \sum_{i=1}^{N_q}  \mathcal{L}_{i,c}
    \end{split}
\end{equation}

For the waiting time in the server queue, the processing time of all previous tasks must be considered. Therefore, based on (\ref{eq: T_i_position}), $T_i$ position in the queue, $\mathcal{L}_{i,q}$ is obtained as, \vspace{-4mm}

\begin{equation}  \label{eq: waiting_time}
    \begin{split}
      \mathcal{L}_{i,q}= \sum_{k=1}^{K_i-1} \sum_{I=1}^{N_a} t_p^I Y^t_{I,k} 
    \end{split}
\end{equation}

Therefore, the sum of all UEs' waiting times in the server is calculated as, \vspace{-5mm}

\begin{equation}  \label{eq: sum_waiting_time}
    \begin{split}
      \mathcal{L}_{q}= \sum_{i=1}^{N_q}  \mathcal{L}_{i,q}
    \end{split}
\end{equation}

By adding both delay types, and normalizing by their maximum values, the total weighted normalized latency is, \vspace{-5mm}

\begin{equation}  \label{eq: normalized_delay}
    \begin{split}
      \mathcal{L}&=\\
      &\frac{\sum\limits_{i=1}^{N_q} \Big(w_i \sum\limits_{j=1}^S j T_s X^t_{i,j}\Big) +\sum\limits_{i=1}^{N_q} \Big( w_i \sum\limits_{k=1}^{K_i-1}  \sum\limits_{I=1}^{N_a} t_p^I Y^t_{I,k} \Big) }{N_a S \times T_s +\sum\limits_{i=1}^{N_a} t_p^i}   
    \end{split}
\end{equation} where $w_i$ is used to prioritize tasks based on their weights. Note that (\ref{eq: normalized_delay}) does not reflect the actual normalized delay of all tasks since some tasks are dropped due to not fulfilling their deadlines. Therefore, Let $d_i$ be the index indicating whether $T_i$ meets its deadline after joint user scheduling and task offloading where $d_i=1$ indicates that the $i^{th}$ task dropped due to missing its deadline. Then, it can be derived as, \vspace{-2mm}

\begin{equation} \label{}
    \begin{split}
        d_i=\begin{cases}
            1, \quad \textit{if } \, \mathcal{L}_i + t_p^i> t_d^i \\
            0, \quad\textit{Otherwise } 
        \end{cases}
    \end{split}
\end{equation} in which $\mathcal{L}_i=\mathcal{L}_{i,q}+\mathcal{L}_{i,c}$ is the $i^{th}$ task total latency. Consequently, by adding the dropped tasks index, (\ref{eq: normalized_delay}) is written as (\ref{eq: normalized_delay_2}).

\begin{figure*}
    \begin{equation}  \label{eq: normalized_delay_2}
    \begin{split}
      \mathcal{L}=\frac{\sum_{i=1}^{N_q} \Big((1-d_i)w_i \sum_{j=1}^S j T_s X^t_{i,j}\Big) +\sum_{i=1}^{N_q} \Big( (1-d_i)w_i \sum_{k=1}^{K_i-1}  \sum_{I=1}^{N_a} t_p^I Y^t_{I,k} \Big) }{N_a S \times T_s +\sum_{i=1}^{N_a} t_p^i}
    \end{split}
\end{equation}
 
\end{figure*}

Since this study involves an \emph{on-off} jammer, the dropped task ratio is not deterministic but a stochastic metric. Therefore, its expected value is considered as the ratio of the expected weighted total number of dropped tasks to the weighted total number of active tasks. It is important to note that a task may be dropped for two reasons: one is the jamming effect, and the other is missing the deadline. Therefore, the expected weighted total number of dropped tasks, $E[N_{\mathbf{D}}]$, can be obtained by subtracting the weighted mini-slot assigned non-jammed processed tasks from the weighted total number of tasks as, \vspace{-3mm}

\begin{equation}  \label{eq: dropped_task_D}
    \begin{split}
      E[N_{\mathbf{D}}]=\sum_{i=1}^{N_a} w_i - \sum_{j=1}^S \sum_{i=1}^{N_a} w_i X^t_{i,j} (1-P^{t}_{i,j}) (1-d_i)
    \end{split}
\end{equation} 
and the dropped task ratio is defined as, \vspace{-3mm}

\begin{equation}  \label{eq: dropped_task_D}
    \begin{split}
      E[\mathcal{D}]&=\frac{E[N_{\mathbf{D}}]}{\sum_{i=1}^{N_a} w_i} \\
      & \hspace{0mm}= \frac{\sum_{i=1}^{N_a} w_i - \sum_{j=1}^S \sum_{i=1}^{N_a} w_i X^t_{i,j} (1-P^{t}_{i,j}) (1-d_i)}{\sum_{i=1}^{N_a} w_i}    
    \end{split}
\end{equation}

Note that by considering the weight of each task, urgent tasks are prioritized in the process.

After deriving the dropped task ratio and latency, the joint task offloading and user scheduling (JTO-US) problem is formulated as follows (17). Note that in (17a), $0 \leq \lambda \leq 1$ is a weight factor to have both latency and dropped task ratio in a multi-objective optimization problem.  \vspace{-2mm}

\begin{equation*}
\hspace{-50mm}\underline{\text{\textbf{Problem: JTO-US}}}
\end{equation*}\vspace{-0.3in}
\begin{mini!}|l|[2]                   
    {X_{i,j}}{ \lambda E[\mathcal{D}] + (1-\lambda) \mathcal{L} \label{eq:min}}{}{}
	\addConstraint{X^t_{i,j}}{\in \{0, 1\},\,  \forall\, i \in \{1,..., N_a\}, \forall \, j \in \{1, ..., S\} \label{eq:X_i_j_1}} 
	\addConstraint{\sum_{i=1}^{N_a}\sum_{j=1}^{S} X^t_{i,j}  }{\leq N_a \, \, \label{eq:X_i_j_2}} 
    \addConstraint{\sum_{i=1}^{N_a} X^t_{i,j} \leq 1 , \quad \forall j=1, ..., S }{ \,\label{eq:X_i_j_3}}    
\end{mini!}

Since $Y^t_{i,k}$ is a non-linear function of $X^t_{i,j}$ (see (\ref{eq: T_i_slot})), \textbf{JTO-US} is classified as a binary non-linear programming problem.  To show the effectiveness of the proposed JTO-US method, genetic algorithm is used as a solver in Section \ref{Sec:Neumerical_results}.

\section{Simulation Results} \label{Sec:Neumerical_results}

\subsection{Parameter Settings}
To explore the effectiveness of the proposed method, simulation is performed and the results are compared with different benchmark algorithms. Tasks are scheduled according to an exponential distribution, with an average arrival rate equals to the duration of one frame to be consistent with the MAC scheduler in 5G. As a result, the stochastic process governing task arrivals is modeled by a Poisson distribution. Tasks are generated independently from one frame to the next, as well as independently across different users. The processing times and the deadline of each task are generated based on the uniform distribution in the interval of $[t_p^{min}, t_p^{max}]$ and $[t_d^{min}, t_d^{max}]$ according to the Table \ref{Table.Param}, respectively.

\begin{table}[!ht]
	\centering
	\caption{Simulation Parameters } \label{Table.Param}
	\begin{tabular}{c|c||c|c} \hline
		\textbf{Parameter}	&	\textbf{Value}  &	\textbf{Parameter}  &  \textbf{Value}  \\
		\hline \hline
		$t_p^{min}$ & $2\, ms $ & $t_d^{min}$ & $5\, ms$ \\ \hline
		$t_p^{max}$ & $10 \, ms$ & $t_d^{max}$ & $50 \, ms $ \\  \hline
		$N_a$ & $10$ & $S$ & $30 $ \\ \hline	
		$T_f$ & $10 \,ms$  & $N_{sim}$ &  $300 $ \\ \hline
	\end{tabular}
\end{table}
To model a jammer behavior, each slot is jammed randomly with a specific probability, $P_{i,j}$ assigned to each user and it is swept from zero to one.

To solve the problem \textbf{JTO-US}, GA is used. The GA settings are listed in Table \ref{Table1.GA_Param}. The other parameters are set empirically and based on initial executions. 
On one hand, a lower number of populations and generations cannot ensure convergence. On the other hand, a very high number of populations and generations results in prolonged convergence time. Additionally, we observe that to prevent the solver from prematurely terminating the process, very low values for \textit{FunctionTolerance} and \textit{ConstraintTolerance} are required\footnote{This is often the case in non-linear binary programming, where many chromosomes across successive generations yield identical or nearly identical values for the objective function. }. 
\begin{table}[!ht]
	\centering
	\caption{Genetic Algorithm Parameters} \label{Table1.GA_Param}
	\begin{tabular}{c|c|c|c} \hline
		\textbf{Parameter}	&	\textbf{Value}  &	\textbf{Parameter}  &  \textbf{Value}  \\
		\hline \hline
		$PopulationSize$ & $400 $ & $ConstraintTo.$ & $10^{-6}$ \\ \hline
		$MaxGen.$ & $500$ & $CrossoverFraction$ & $0.5$ \\  \hline
		$FunctionTol.$ & $10^{-10}$ & $MaxStallGen.$ & $50$ \\ \hline	
	\end{tabular} 
\end{table}

\subsection{Performance Evaluation}
Two main scenarios are considered in the simulation. 
In the first scenario, we treat all the users the same by setting the weights equal to one. Tasks are generated at the beginning of each frame randomly and the mini-slots are jammed independently and according to a certain probability. The jamming probability is swept from zero to one and the results are collected and then averaged over all $300$ simulation runs. GA is used as a solver\footnote{The surrogate algorithm is also used as a solver; however, better performance for GA is seen.} For the performance evaluation, three other algorithms are used as benchmark, namely, $JTO-US_{_{GA-Nj}}$ (Genetic algorithm without considering the jamming effect), SJF (shortest job first), and SDF (shortest deadline first).

\begin{figure}[htp]
    \centering
    \includegraphics[width=.8\linewidth]{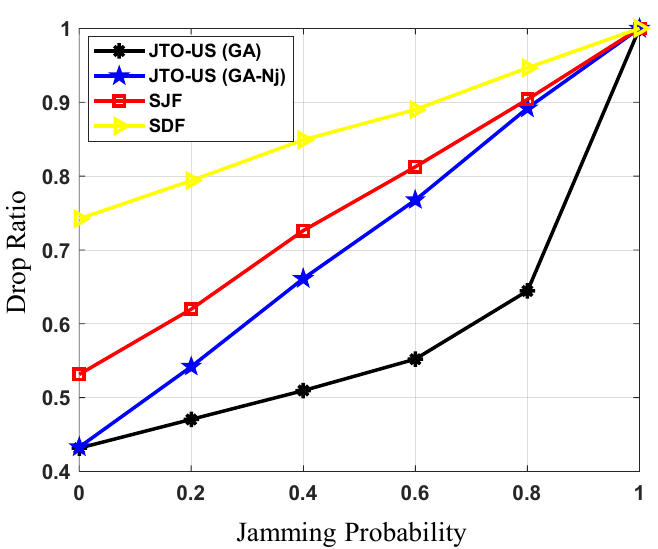}
    \caption{Drop ratio vs. jamming probability.}
    \label{fig_drop_ratio}
 \end{figure}

Fig. \ref{fig_drop_ratio} illustrates the drop ratio as a function of jamming probability. Overall, the proposed method ($JTO-US_{_{GA}}$) solved by GA exhibits the lowest drop ratio, followed by the $JTO-US_{_{GA-Nj}}$ algorithm, with SJF and SDF having higher drop ratios. However when $P_{i,j}=0$, GA and GA-Nj both have the same drop ratio lower than the other two. This is intuitive since in this case, there is no jammer in the environment so $JTO-US_{_{GA-Nj}}$ turns to act the same as $JTO-US_{_{GA}}$. In this case, the drop ratio is related to the tasks that are not  assigned any mini-slot or their deadline is already expired. It is clear that when $P_{i,j}=1$, all algorithms have the same drop ratio of one since all tasks are droped due to jamming effect.

\graphicspath{{figs/}}

\begin{figure}[htp]
    \centering
    \includegraphics[width=.8\linewidth]{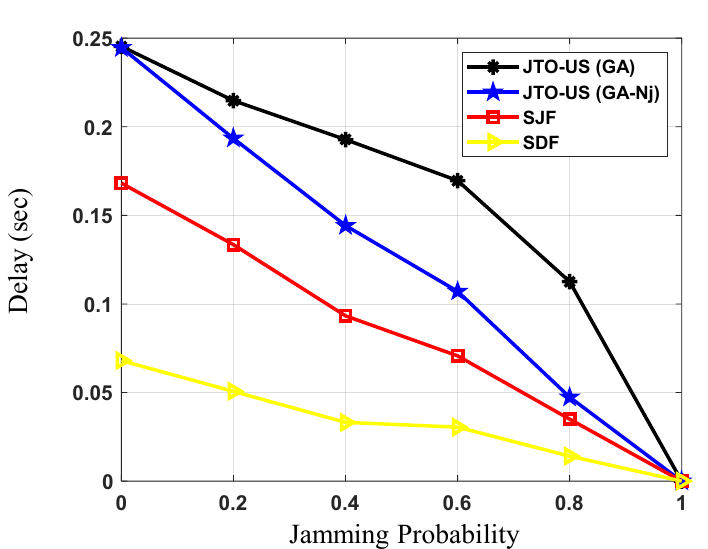}
    \caption{Delay vs. jamming probability.}
    \label{fig_delay}
 \end{figure}

The average latency versus jamming probability is plotted in Fig. \ref{fig_delay}. It is observed that SDF has the lowest latency followed by SJF, and then $JTO-US_{_{GA-Nj}}$ and $JTO-US_{_{GA}}$, respectively. Indeed, the latency follows the opposite pattern of the drop ratio. This is because the lower the drop ratio, the higher the number of tasks in the MEC server queue, and hence, the longer the expected waiting time.

\graphicspath{{figs/}}

\begin{figure*}
    \centering
    \begin{subfigure}[b]{0.3\textwidth}
        \includegraphics[width=\textwidth]{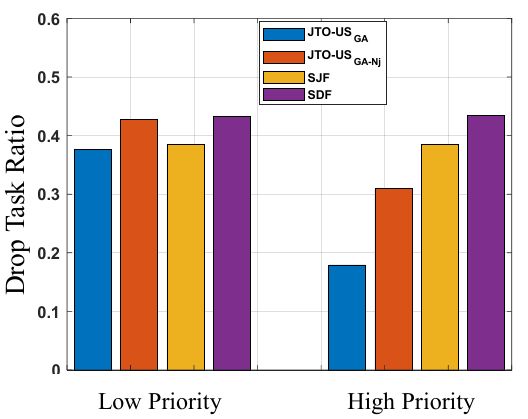}
        \caption{Distribution of the dropped tasks vs. Tasks' weights- $P_{i,j}=0.1$.}
        \label{fig: fig_dis_weight_1}
    \end{subfigure}
    ~ 
    \begin{subfigure}[b]{0.3\textwidth}
        \includegraphics[width=\textwidth]{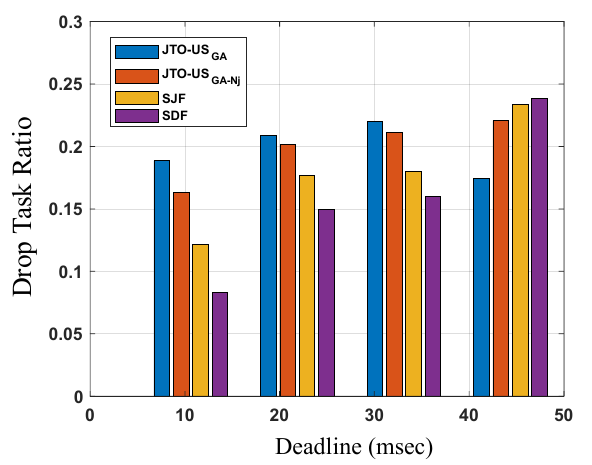}
        \caption{Distribution of the dropped tasks vs. Tasks' deadline- $P_{i,j}=0.1$.}
        \label{fig: fig_dis_deadline}
    \end{subfigure}
    ~ 
    \begin{subfigure}[b]{0.3\textwidth}
        \includegraphics[width=\textwidth]{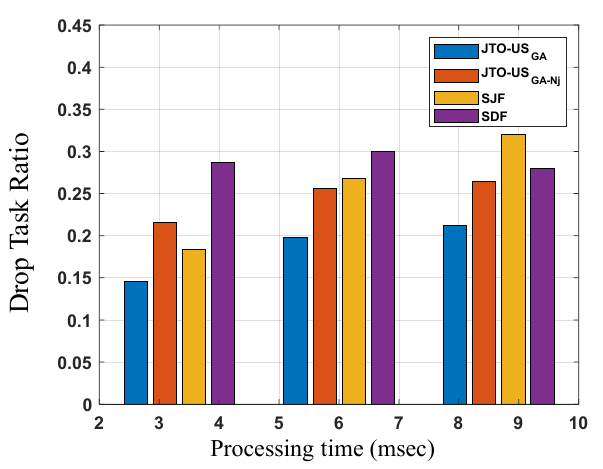}
        \caption{Distribution of the dropped tasks vs. Tasks' processing time- $P_{i,j}=0.1$.}
        \label{fig:fig_dis_process}
    \end{subfigure}
    \caption{Distributions of the dropped tasks $P_{i,j}=0.1$}\label{fig:fig_dist_1}
\end{figure*}

In the second scenario, we used tasks with two different priorities. For low-priority tasks, $\omega_i=0.1$ and for high-priority ones $\omega_i=0.9$\footnote{Low- and high-priority tasks are evenly generated in each mini-slot, with a probability of 0.5 for each to have a fair comparison when exploring the distribution of the dropped tasks.} Then the distribution of the dropped tasks is extracted based on the tasks' properties\footnote{In deriving such a distribution, we neglect the weights in calculating drop task ratio to observe the effect of priority on the dropped tasks.}. Fig. \ref{fig: fig_dis_weight_1} shows the drop task ratio for each category of low and high priority. It is observed that the proposed algorithm, $JTO-US_{_{GA}}$, effectively manages priority tasks, as the ratio for high-priority tasks (0.18) is significantly lower than that of low-priority tasks (0.37). $JTO-US_{_{GA-Nj}}$ also performs well in dealing with high and low priority tasks. However, the drop ratio is higher, specifically for high-priority tasks, since it does not take the jamming behaviour into account. The other two, SDF and SJF, do not show any significant attempt at handling tasks with priority. Note that the jamming probability in this experiment is set to 0.1.

\begin{figure*}
    \centering
    \begin{subfigure}[b]{0.3\textwidth}
        \includegraphics[width=\textwidth]{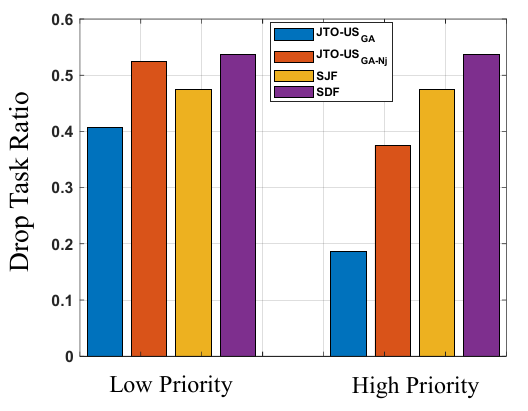}
        \caption{Distribution of the dropped tasks vs. Tasks' weights- $P_{i,j}=0.6$.}
        \label{fig:fig_dis_weight_2}
    \end{subfigure}
    ~ 
    \begin{subfigure}[b]{0.3\textwidth}
        \includegraphics[width=\textwidth]{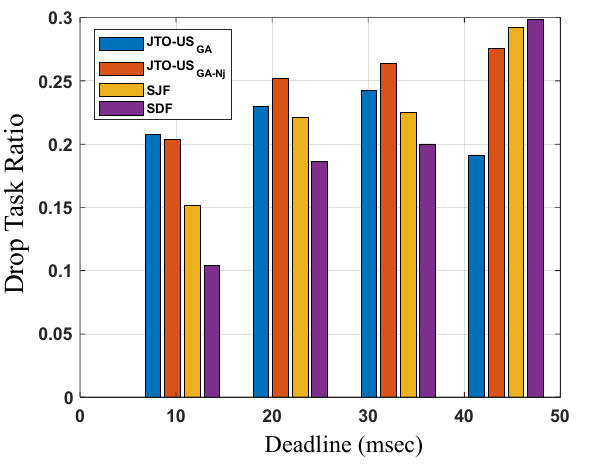}
        \caption{Distribution of the dropped tasks vs. Tasks' deadline- $P_{i,j}=0.6$.}
        \label{fig:fig_dis_deadline_2}
    \end{subfigure}
    ~ 
    \begin{subfigure}[b]{0.3\textwidth}
        \includegraphics[width=\textwidth]{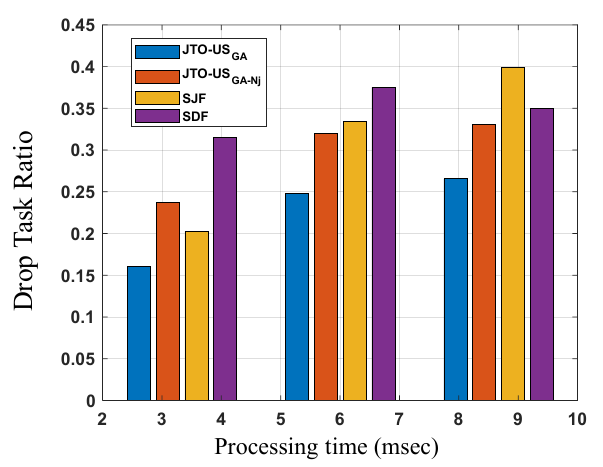}
        \caption{Distribution of the dropped tasks vs. Tasks' processing time- - $P_{i,j}=0.6$.}
        \label{fig:fig_dis_process_2}
    \end{subfigure}
    \caption{Distributions of the dropped tasks $P_{i,j}=0.6$}\label{fig:fig_dist_2}
\end{figure*}

The distribution of the dropped tasks based on tasks' deadlines is plotted in Fig. \ref{fig: fig_dis_deadline}. It is notable that SDF has the lowest number of dropped tasks for tasks with shorter deadlines. As the deadlines of the tasks increases from 5 msec to 50 msec, the drop task ratio increases from 0.06 to 0.24. However, total drop task ratio in SDF (0.84) is higher than that of the proposed method (0.55).

The distribution of dropped tasks based on processing time is shown in Fig. \ref{fig:fig_dis_process}. Although it is expected that SJF has the lowest drop ratio for tasks with shorter processing times, $JTO-US_{_{GA}}$ shows better performance (0.146 compared to 0.18). As the processing time increases from 1 ms to 10 ms, the dropped task ratio rises for both $JTO-US_{_{GA}}$ and SJF, maintaining the $JTO-US_{_{GA}}$ the lowest among all methods. 

The same simulation scenario but with jamming probability $P_{i,j}=0.6$ is conducted and the results are plotted in Figs. \ref{fig:fig_dis_weight_2}, \ref{fig:fig_dis_deadline_2}, and \ref{fig:fig_dis_process_2}. An increase in $P_{i,j}$ further accentuates the difference in dropped task ratio between the proposed method and others. As expected, the ratio rises for all methods as the jamming probability increases. However, the ratio for $JTO-US_{{GA}}$ remains at a reasonable level of 0.6, whereas it exceeds 0.9 for the other methods. Specifically, Fig. \ref{fig:fig_dis_weight_2} illustrates that the proposed method achieves drop ratios of 0.19 for low-priority tasks and 0.41 for high-priority tasks. In contrast, for $JTO-US{_{GA-Nj}}$, the drop ratio increases to 0.53 and 0.38 for low- and high-priority tasks, respectively.

\section{Conclusions}\label{Section conclusion}
We have proposed a novel joint task offloading and user scheduling (JTO-US) algorithm to address the challenges posed by jamming attacks in 5G mobile edge computing (MEC) systems. By leveraging a genetic algorithm (GA) for optimization, our approach successfully minimizes both the task delay and the dropped task ratio. The proposed $JTO-US_{_{GA}}$ framework outperforms benchmark scheduling methods by incorporating the proposed dynamic jamming mitigation into the scheduling-offloading process. Our simulation results demonstrate that $JTO-US_{_{GA}}$ achieves superior performance, particularly under severe jamming conditions. For instance, when the jamming probability reaches 0.8, the drop ratio is reduced to $63\%$, compared to $89\%$ with the next best method. Moreover, the algorithm effectively prioritizes high-priority tasks, ensuring that critical tasks are handled with minimal interruption, further improving system reliability and QoS. The findings of this study highlight the effectiveness of joint optimization in handling jamming attacks and improving overall system performance. Future work will extend the proposed model to multi-server environments and explore the integration of diversity transmissions to enhance reliability for ultra-reliable low-latency communication (URLLC) scenarios.

\section*{Acknowledgment}

This work was supported in part by funding from the Innovation for Defence Excellence and Security (IDEaS) program from the Department of National Defence (DND) and the Natural Science and Engineering Research Council (NSERC) of Canada CREATE TRAVERSAL Program

\bibliographystyle{IEEEtran}

\end{document}